\documentstyle{article}

\title{Temporally Asymmetric Fluctuations are
 Sufficient for the Operation of a Correlation Ratchet}
\author{Dante R. Chialvo$^2$  and  Mark M. Millonas$^{1,2}$\\
\small $\ ^1$Complex Systems Group, Theoretical Division,  and Center for
Nonlinear Studies,\\ \small MS B258, Los Alamos National
Laboratory, Los Alamos, NM 87545, USA.\\
\small    $^2$Fluctuations and Biophysics Group, Santa Fe Institute,\\ \small
1660 Old
Pecos Trail, Suite A, Santa Fe, NM 87501,  U.S.A.\\  }

\begin{document}

\maketitle

{\bf
  A number of recent attempts to understand
 broad principles of energy transduction in biological
systems have focused on correlation ratchets---systems
which extract work out of  fluctuations which are correlated in
time.\cite{mag,me1,ast,prost,doering,oster}
 Correlation ratchets  are  ``information engines"
analogous to Maxwell's Demon,  which extract work out of a  bath
by  using information about the system to ``choose" only those
fluctuations which are helpful to make the engine run.\cite{me2}
This information, which can only be acquired if the Demon is not in equilibrium
with the bath,\cite{B} can be   used to rectify the energy already available,
 but otherwise inaccessible in the thermal bath.
Processes like this,  in which
the energy stored in a nonequilibrium bath
 is transformed into work at the expense of increased entropy,
are believed to be the basis of ``molecular motors", and are of great
importance in biology, and a number of other fields.
It has been shown
that the combination of a broken spatial symmetry in the potential (or ratchet
potential) and time correlations in the driving are crucial, and enough to
allow
the transformation of the fluctuations into work.\cite{mag}
The required broken spatial
symmetry implies a specific molecular arrangement of the proteins involved.
Here we show that a broken spatial symmetry is  {\sl not required}, and that
 temporally asymmetric fluctuations (with mean zero)  can be used to do work,
{\sl even when the ratchet potential is completely symmetric}.
Temporal asymmetry, defined as a lack of invariance of the statistical
properties under the operation to temporal inversion,  is a generic property of
nonequilibrium fluctuation, and should therefore be expected to be quite common
in biological systems.}

\vspace{.5in}

In biological systems
energy stored in the form of ATP molecules is used to drive various biochemical
processes.  Since new ATP molecules are constantly being made, and the degraded
ones removed from the system, the bath of energy molecules is very far from
equilibrium.  The question of how this tremendous negative
entropy allows for an effective transduction of energy at the nanoscale in a
very noisy environment is greatly complicated by the fact that there are no
principles of the power and generality of those of equilibrium statistical
mechanics for nonequilibrium systems.

It has been found that, under the right
conditions spatially homogenous fluctuations with  mean zero can be used to do
work if they are correlated in time.\cite{mag,me1,ast,prost,doering,oster}
These systems we call ``correlation ratchets".\cite{nn1}  These time
correlation
are  due to the coupling of a system to a bath which is not in equilibrium, and
patently do not violate the second law of thermodynamics.   These systems  are
indeed analogs of Maxwell's Demon, and operate as information engines.  As
shown
by Szilard\cite{Szilard}, the information is acquired at the expense of an
entropy increase of the Demon, an  observation which salvages the second law.
Similarly it has been shown that correlation ratchets works  at the expense of
the total increase of entropy of the bath, and run off of the physical
information (negentropy) contained in the nonequilibrium bath.\cite{me2}

Here we consider the simplest imaginable system which contains the crucial
elements.   We strongly emphasize here that it is not the model, which is
indeed
trivial, or its simplifying assumptions,  but the physical principal of
temporal asymmetry we illustrate which is important.  This physical principle
has
not been discussed before to the best of our knowledge in any context, and
certainly not with respect to correlation ratchets and biological energy
transduction.  Although the specific  model has been  chosen as a simple
illustration, understandable to a broad audience, the behavior which it
exhibits
is quite general indeed, and we hope extension to more complicated and
realistic
situation will present themselves naturally to the mind of any
attentive reader who grasps our main point.

 We consider a particle in the
piecewise linear potential pictured in Fig. 1(a), as considered in \cite{mag}.
The potential is periodic and extends to infinity in both directions. $\lambda$
measures the spacing of the wells, $\lambda_1$ and $\lambda_2$ the inverse
steepnesses of the potential in opposite directions out of the wells, and $Q$
the
well depths. The particle undergoes overdamped Brownian motion due to its
coupling with a thermal bath of temperature $T$, and external driving $F(t)$
with represent the nonequilibrium forces.   The  expression for the current in
the adiabatic limit, which measures the work done by the ratchet has already
been derived  in this  case,\cite{mag,ris} where \begin{equation} J(F) =
{P_2^2    \sinh(\lambda    F/2 kT)\over kT  (\lambda/Q)^2 P_3 - (\lambda/Q)P_1
P_2 \sinh(\lambda F/2kT)} \end{equation} \begin{equation} P_1 = \delta +
{\lambda^2 -\delta^2\over      4}{F\over  Q},\ \ \ P_2=  \left[ 1-{\delta
F\over   2 Q}  \right]^2   - \left( {\lambda   F\over   2  Q} \right)^2
\end{equation} \begin{equation} P_3 =   \cosh[(Q- \delta F/2)/kT]  -
\cosh(\lambda  F/2kT)
      \end{equation}
where   $\lambda = \lambda_1 + \lambda_2$ and  $\delta =  \lambda_1-\lambda_2$.
The  average  current,  the   quantity  of primary interest, is   given  by
\begin{equation}
<J> =   {1\over \tau} \int_0^\tau J(F(t))\  dt
\end{equation}
where  $\tau$  is  the  period    of   the  driving   force $F(t)$,  which  is
assumed longer than  any other   time  scale   of the     system   in
this    adiabatic limit.
 Magnasco considered  this case, but only for
$F(t)$ symmetric  in time  $F(t)
=F(n\tau-t)$.  Here    we    will  again  consider a  driving  with
a   {\it zero mean}, $<F(t)> \ =0$, but  which is asymmetric in time,
\begin{equation}
F(t) = \left\{ \begin{array}{lll}
\left({1+\epsilon\over 1-\epsilon}\right)A & \ \ \   & 0 \leq
  t< {\tau\over 2}(1
-\epsilon),\ \ mod \ \tau \\
 - A & \ \ \   & {\tau\over 2}(1-\epsilon) > t \leq \tau,\ \ \ \ mod \ \tau
\end{array}\right.
  \end{equation}
as shown in
Fig. 1(b).
  In this case the  time averaged current  is  easily calculated,
\begin{equation}
<J> =   {1\over 2} \left[(1+\epsilon) J(A)  + (1-\epsilon)J(-(1+\epsilon)
A/(1-\epsilon))\right] .    \end{equation}
Fig. 2 shows that
the current is a peaked function of the amplitude of the driving.  Thus,
everything else being constant, there is an optimal amplitude for the driving.
Similarly Fig. 3 shows that the current is also a peaked function of $kT$.  The
driving, the potential, and the thermal noise in fact play
 cooperative roles. Unless $A$ is quite large, there are no transition out
 of the wells when $kT = 0$, and therefore no current, but if the noise is too
large it washes out both the potential and the details of the driving, and the
current again goes to zero.  Similarly, without the driving the transitions in
either direction are the same, but if the driving is too large the potential
plays less of a selective role, and the current drops back down. Here the
 main features introduced by the temporal asymmetry are the interplay of the
lower potential barriers in the positive direction relative to the negative
direction (for this particular driving) and the corresponding shorter and
longer times respectively the force is felt.  These types of competitive
effects
appear ubiquitously in systems where there is an interplay between thermal
activation and dynamics, including, but by no means limited to,  the phenomenon
of stochastic resonance.

 Figs. 2 and 3 are for completely symmetric potentials. In these cases
the exponential Arrhennius dependence of the thermal activations over the
barriers overcomes the time factors, and the current is in the opposite
direction of that which is produced by a temporal symmetry and a spatial
asymmetry. This effect is of course reversed if $\epsilon \to -\epsilon$. There
can also be competitive effects between the temporal asymmetry and the
spatial asymmetry, as pictured in Fig. 4, which gives rise to a very unusual
switching of the direction of the current as the asymmetry factor $\epsilon$ is
varied.  This reversal represent the competition of the spatial asymmetry,
which dominates for small $\epsilon$ an the temporal asymmetry, which dominates
for large $\epsilon$.

Temporal asymmetry is definitely {\it not}
related to the simplified  adiabatic approximation presented above, and is just
as valid for fast noise.   For instance,  the general explanation of the
phenomena shown in Figs. 1-3 given above  is equally applicable to the case
where the period of the driving is large with respect to the intra-well
relaxation time, but not large with respect  to the first-passage times over
the
barriers, such it as the case of stochastic resonance.\cite{nnn}   Additional
argument can be made for even faster noise.

 In cannot be emphasized enough that the reasons
for the phenomena we discuss here are not related to the specific
approximations
which allowed for an analysis solution in our very simple illustration, but are
ubiquitous characteristics of nonequilibrium fluctuations, and deserve to be
studied in more detail.  The main idea here, that of a temporal
asymmetry in the driving, can be easily incorporated by extending the theory of
Ref. \cite{me1} to continuous, but non-Gaussian noise,\cite{el}
 and
the discrete noise models of Ref. \cite{doering}, with qualitatively similar
results.    In addition, fluctuating potential  models\cite{ast,prost,oster}
can
be made to in corporate temporal asymmetry as an important factor if the
potential
changes shape as it fluctuated in magnitude.  A detail example of this type of
system is also forthcoming in a separate publication.\cite{el2}

Our
main point here, which should be clearly understood, is that {\it temporal
asymmetry, defined as a lack of invariance of the statistical properties of a
system under the operation of time inversion,  can be expected to be ubiquitous
property of most nonequilibrium systems, and can give rise to nonequilibrium
transport, such as that which is utilized by many biological processes}.  We
believe it is a  fundamental principle, equal
in importance to Magnasco's recent observation that spatial asymmetry and
temporal correlations are sufficient for nonequilibrium transport.\cite{mag}

This
principle has never been pointed out before, even though it might easily  have
been demonstrated by means of rather simple extensions of the   models
of Ref. \cite{mag,doering} if the respective authors had realized the
general significance of this type of nonequilibrium asymmetry. The
authors of Ref.  \cite{doering} came close to this point when they pointed out
that an spatially asymmetric marginal density of the dichotomous fluctuations
they use would give rise to a current.  In their specific context this might
under certain conditions be regarded as a type of temporal asymmetry, but they
neither analyze this case, nor draw any distinction between this situation
and the case where there is a net force.  More importantly, the more general
conclusions we draw here cannot be inferred from this comment, which only
applies
to their specific model.  They do not point out, as we do here for the first
time, that  temporal asymmetry is the crucial property, and the this property
is
a quite generic aspect of nonequilibrium fluctuation, and not  a property of a
specific model.

Time correlations in the driving will influence the
transitions in either direction.  Since this influence depends on the shape of
the potential, as well as the properties of the noise, an asymmetry in the
potential will give rise to a net current.  However, when the noise is
temporally asymmetric, its correlation properties {\it in either direction are
different, and a net current can arise even in the absence of a spatial
asymmetry.}  Note that the dependence of the strength and direction of the
current on the properties of the noise discussed here is not related to the
similar results of Refs. \cite{me1} and \cite{doering}, in  which only
temporally symmetric fluctuations where considered, and the current vanished in
the spatially symmetric case.

Such temporal asymmetry is present in virtually any waveform or noise with a
nontrivial distribution of frequency component phases.  It
is thus interesting to note that this possibility has not  been considered
explicitly in any previous study of correlation ratchets.  It is believed that
many driven biochemical processes work by cycling though a sequence of
intermediary states.  These cycles are driven by the steps in the hydrolysis of
ATP.  If the steps in this sequence have  different time scales, which is
generally the case, the result will be temporal asymmetry, as well as
correlations.  Unless there is a physical (generally equilibrium) restriction,
temporal asymmetry will probably be ubiquitous in nonequilibrium systems, such
as biological energy transducers.

It is exciting to note that now
that the motion of individual motor proteins can be tracked at the level of
single steps\cite{sv,sv2}, it may be possible to test some of these ideas via
an
analysis of the time series of the motion of the molecule.  Temporal asymmetry
will appear as the nonvanishing of the odd moments (of any order other than
first, e.g. the bicovariance $<F(t) F^2(s)>$) of the fluctuations, an
observation which may be of some use to experimentalists.

\noindent ACKNOWLEDGMENTS: Work partially supported by NIMH. The Santa
Fe  Institute  receive funding from the John D. Catherine T. MacArthur
Foundation, NSF, and the U.S. Department of Energy.

\begin{figure}[t]
\vspace{4in}
\caption{\sf (a) Plot of the simple piecewise ratchet potential, where the
spatial degree of asymmetry is given by the parameter $\delta=
\lambda_1-\lambda_2$.  (b) Plot of the driving force $F(t)$ which preserved
zero mean $<F(t)>\ =0$, where the temporal asymmetry is given by the parameter
$\epsilon$.  }  \end{figure}

\begin{figure}[t]
\vspace{4in}
\caption{\sf Plot of the current vs. $A$ for a symmetric potential
($\delta = 0$), with $Q=1$,$\lambda=1$, $\epsilon = 1/2$ and (a) $kT = .3$,
(b) $kT= .1$, (c) $kT=.01$.}  \end{figure}

\begin{figure}[t]
\vspace{4in}
\caption{\sf
Plot of the current vs. $kT$ for a symmetric potential ($\delta =
0$), with $Q=1$,$\lambda=1$, $\epsilon = 1/2$ and (a) $A = 1$, (b) $A= .8$,
(c) $A=.5$.}
\end{figure}

\begin{figure}[t]
\vspace{4in}
\caption{\sf Plot of the current vs. the time asymetry parameter $\epsilon$
 with $Q=1$,$\lambda=1$,$kT = .01$, and $A=2.1$, for (a) a symetric
potential $\delta =0$  and  asymmetric potentials (b) $\delta = -.3$ (c)
$\delta = -.7$}
\end{figure}


\begin{thebibliography}{99}
\bibitem{nn1} This term appears to have been introduced by George Oster in
regards to fluctuating potential models, but here we extend this terminology to
cover fluctuating force modes as well.
\bibitem{nnn}  It should be noted here  that  for the same reason the basic
phenomena described in  Magnasco's illustration (Ref. [1]) will also be more or
less unchanged in this faster noise case.  Something which seems to have
escaped a number of authors.
\end{thebibliography}
\end{document}